\documentclass[a4paper]{jpconf}
\usepackage{graphicx}

\begin{document}

\title{Spin effects in strong-field laser-electron interactions}
\author{S Ahrens$^1$, T-O M\"uller$^{2}$, S Villalba-Ch\'avez$^3$, H Bauke$^1$ and C~M\"uller$^{3}$}

\address{$^1$ Max-Planck-Institut f\"ur Kernphysik, Saupfercheckweg 1, 69117 Heidelberg, Germany}
\address{$^2$ Physik Department, Technische Universit\"at M\"unchen, 85747 Garching, Germany}
\address{$^3$ Institut f\"ur Theoretische Physik I, Heinrich-Heine-Universit\"at, 40225 D\"usseldorf, Germany}

\ead{c.mueller@tp1.uni-duesseldorf.de}


\begin{abstract}
The electron spin degree of freedom can play a significant role in relativistic scattering processes involving intense laser fields. 
In this contribution we discuss the influence of the electron spin on (i) Kapitza-Dirac scattering in an x-ray laser field of high intensity, (ii) photo-induced electron-positron pair production in a strong laser wave and (iii) multiphoton electron-positron pair production on an atomic nucleus.
We show that in all cases under consideration the electron spin can have a characteristic impact on the process properties and their total probabilities.
To this end, spin-resolved calculations based on the Dirac equation in the presence of an intense laser field are performed. The predictions from Dirac theory are also compared with the corresponding results from the Klein-Gordon equation.

\end{abstract}

\section{Introduction}
Quantum mechanical interaction processes between electrons and high-intensity laser fields in a relativistic regime have encountered a sustained interest in recent years (see \cite{review1,review2} for reviews). A particular question in this regard is to which extent the electronic spin degree of freedom may influence the electron-laser interaction. When a free electron is exposed to the electromagnetic field of a plane laser wave, the electronic spin vector performs a precessive motion. However, after the laser pulse has passed and the interaction is over, the electron will be in the same spin state as it was before the interaction started \cite{Walser}.

Distinct electron spin transitions can occur when the electron participates in a scattering process while being subject to the laser field. Then, the electronic spin state before and after the interaction with the field may change. Corresponding spin transitions have theoretically been revealed, for instance, in multiphoton Compton scattering \cite{PanekCom,SerboCom,Boca}, laser-assisted Mott scattering \cite{Szymanowski,PanekMott}, and photo-induced production of electron-positron pairs \cite{Tsai,SerboPP}. Spin effects in strong laser fields were also investigated with regard to bound-electron dynamics \cite{Hu}, two-photon ionization \cite{Faisal} and high-harmonic generation \cite{HHG}. Moreover, characteristic differences between fermionic and bosonic particles were found in pair production in an oscillating electric field \cite{Popov} and in recent investigations of the Klein paradox \cite{Krekora,Wagner}.

In this contribution, further strong-field laser-electron interaction processes are studied with respect to the relevance of spin effects. First, we consider Kapitza-Dirac scattering of electrons on a standing laser wave of high intensity and frequency (Section 2). Our analysis relies on the Dirac equation which is solved by numerical integration. In a certain regime of interaction parameters, characteristic electron spin dynamics involving Rabi oscillations and complete spin-flip transitions is identified. Differences between the Kapitza-Dirac scattering of spin-half and spin-zero particles are emphasized. Afterwards, we turn to spin-sensitive calculations of electron-positron pair creation in very intense laser fields. By using the polarization operator in the presence of a plane-wave laser field, we examine the production of pairs induced by a high-energy photon travelling through a plane laser wave of circular polarization (Section 3). A comparison between the total production rates of spin-half versus 
spin-zero particles is drawn, this way complementing the earlier studies in \cite{Tsai,SerboPP}. Finally, electron-positron pair production in combined laser and nuclear Coulomb fields is studied within an $S$-matrix formalism using Dirac-Volkov states and the spin projection operator (Section 4). In a nonperturbative regime of interaction, the degree of longitudinal polarization of one of the produced particles is obtained and its dependence on the particle momentum is discussed. We finish with some concluding remarks (Section 5).

Relativistic units with $\hbar=c=4\pi\varepsilon_0=1$ are used unless explicitly stated otherwise.

\section{Spin dynamics in Kapitza-Dirac scattering}
The quantum mechanical diffractive scattering of an electron beam from a standing wave of light, formed by two counterpropagating waves of equal frequency and intensity, is referred to as Kapitza-Dirac effect \cite{Batelaan}. The process, which relies on the quantum wave nature of the electron, constitutes an analogue of the classical diffraction of light on a grating, but with the roles of light and matter interchanged. In its original version  \cite{KD}, the effect represents a combined absorption and emission process involving two photons: the electron absorbs one photon of momentum ${\vec k}$ from, say, the right-travelling laser beam and emits a photon of momentum $-{\vec k}$ into the counterpropagating laser beam. The momentum of the incident electron thus changes by $2{\vec k}$. Seminal theoretical work on the Kapitza-Dirac effect and its multiphoton generalization to intense laser fields has been carried out starting in the 1960s \cite{Fedorov}. 

It is interesting to note that a clear experimental confirmation of the Kapitza-Dirac effect as originally proposed has been achieved only recently \cite{Freimund}. This observation has led to a newly revived interest of theoreticians in the effect. During the last decade, a variety of additional aspects of Kapitza-Dirac scattering has been studied such as the influence of the electronic wave packet size \cite{Efremov} and diffractive scattering in the case of counterpropagating laser beams of non-equal frequencies \cite{Smirnova}. Besides, an analysis based on the Pauli equation showed that spin effects are very small at low laser intensities \cite{Rosenberg}. Both the experiment and the theoretical studies have relied on rather moderate, nonrelativistic laser intensities \cite{Avetissian}.

The recent advent of novel high-intensity light sources raises the question as to how the Kapitza-Dirac effect is modified in the relativistic regime where also spin effects may be expected to occur. To this end, a treatment of the process based on the Dirac equation is required. It has been provided recently \cite{Sven}. One can show that -- rather than in the original Kapitza-Dirac effect with two-photon exchange -- pronounced spin effects may arise when the electron scatters from the standing wave with participation of an {\it odd} number of laser photons. Such an interaction channel can be accessed by a suitable choice of the electron and field parameters. The latter can be determined from a generalized Bragg condition which follows from the laws of energy and momentum conservation. 

The relevant Dirac equation may be written as
\begin{equation}
  \label{eq:Dirac}
  i \frac{\partial }{\partial t} \Psi({\vec x},t)
  = \left[
    \left( 
      \hat{\vec p} + e\vec A (\vec x, t)
    \right)\cdot \vec \alpha + \beta m
  \right]\Psi(\vec x,t)\,,
\end{equation}
where $-e$ and $m$ are the electron charge and mass, respectively, $\hat{\vec p}=-i\nabla$ is the momentum operator, and $\vec A (\vec x, t)$ denotes the vector potential of the standing laser wave. Besides, $\vec\alpha=(\alpha_x,\alpha_y,\alpha_z)$ and $\beta$ are the Dirac matrices. Due to the periodicity of the vector potential, the electronic wave function may be decomposed into a discrete set of plane waves, according to
\begin{equation}
  \Psi({\vec x},t) = 
  \sum_{n,\zeta} c_n^{\zeta}(t) \psi_{n,\vec p}^\zeta (\vec
  x)\,,\quad 
  \label{eq:plane-wave-ansatz} 
\end{equation}
with free Dirac states $\psi_{n,\vec p}^\zeta (\vec x)=u_{\vec p+n\vec k}^{\zeta}\, \e^{i(\vec p + n \vec k)\cdot \vec x}$ of momentum $\vec p+n\vec k$ ($n=0,\pm1, \pm2,\dots$).
The index $\zeta \in \mbox{$\{ {+}\uparrow , {+}\downarrow, {-}\uparrow, {-}\downarrow \}$}$ labels the sign of the energy and the spin projection along the laser electric field vector.
Insertion of the ansatz (\ref{eq:plane-wave-ansatz}) into the Dirac equation (\ref{eq:Dirac}) yields a system of coupled differential equations for the expansion coefficients $c_n^{\zeta}(t)$, which represent the transition amplitudes to the various states. The system of equations can be solved by numerical integration, starting from the initial condition $c_0^{+\uparrow}(0)=1$ and all other $c_n^\zeta(0)=0$. It describes an incident electron of momentum $\vec p$ and positive spin projection.

As an example, we consider Kapitza-Dirac scattering of an electron which is incident with a weakly relativistic momentum of 176 keV at a small angle of $\vartheta=0.4^\circ$ with respect to the laser beam axis. The laser frequency amounts to 3.1 keV and each laser beam has a peak intensity of $2.0\times10^{23}$\,W/cm$^2$. Under these circumstances, the electron may be scattered with absorption of two photons from the right-travelling laser beam and emission of one photon into the left-travelling beam. The electron momentum thus changes by $3{\vec k}$.

\begin{figure}[h]
\includegraphics[width=22pc]{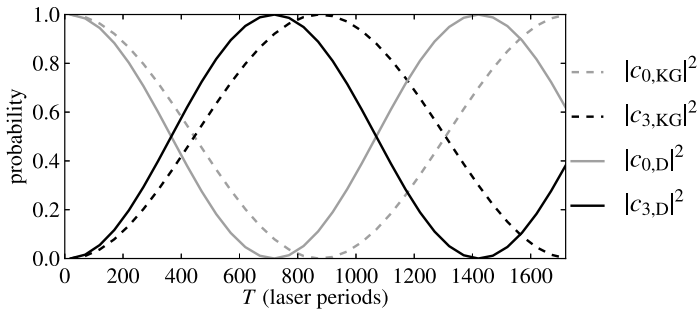}\hspace{2pc}%
\begin{minipage}[b]{14pc}\caption{\label{Rabi} Rabi oscillations of the occupation probabilities $|c_0|^2$ and $|c_3|^2$ of the initial state and the scattered state, respectively, as a function of the interaction time $T$ for a three-photon Kapitza-Dirac effect. The Rabi flopping dynamics is faster for a Dirac (D) particle than a Klein-Gordon (KG) particle.}
\end{minipage}
\end{figure}

Figure \ref{Rabi} shows the time dependence of the occupation probabilities of the initial state (with momentum $\vec p$) and the scattered state (with momentum ${\vec p}+3{\vec k}$). The latter probability consists of two contributions, corresponding to a spin-preserving or spin-flipping transition. In the figure, these contributions have been added to give the spin-summed probability $|c_3(T)|^2:=|c_3^{+\uparrow}(T)|^2 + |c_3^{+\downarrow}(T)|^2$ (see the black solid line) \cite{negative}. Note that the results shown in Fig.\,\ref{Rabi} represent occupation probabilities {\it outside} the laser field; that is, after each value of $T$ the interaction between the electron and the field has been faded out in the numerical simulation.

The occupation probability oscillates back and forth between the two momentum modes ${\vec p}$ and ${\vec p}+3{\vec k}$, with a Rabi frequency of \cite{Sven}
\begin{equation}
  \label{eq:Rabi_DE}
  \Omega_{R,{\rm D}} = \Omega_0
  \sqrt{\frac{25}{2}\left(
      \frac{p_E}{k}
    \right)^2 + 1}\,.
\end{equation}
Here, we introduced the abbreviation $\Omega_0={e^3E^3}/({24m^3k^2})$, with the laser electric field amplitude $E$, and the electron momentum component $p_E=|\vec p\,|\sin\vartheta$ along the electric field.

It is important to note that, if the electron was a spinless particle, the Rabi frequency would differ from Eq.\,(\ref{eq:Rabi_DE}). For a Klein-Gordon particle it would read
\begin{equation}
  \label{eq:Rabi_KGE}
  \Omega_{R,\,\rm{KG}} = 
  \Omega_0\frac{5}{\sqrt{2}}\frac{p_E}{k}\,.
\end{equation}
The difference may be understood by the following argument. In order to induce a transition involving three photons, it is necessary to have  in the interaction Hamiltonian terms which are linear in the laser field. In the case of a spinless particle, the $\hat{\vec p}\cdot{\vec A}$ interaction term is of this kind. It leads to the linear momentum dependence of the Rabi frequency in Eq.\,(\ref{eq:Rabi_KGE}). In the case of a spin-half particle there is in addition a term which corresponds to the magnetic ${\vec \sigma}\cdot{\vec B}$ interaction in the Pauli equation. This term, which is momentum-independent and capable of inducing spin-flip transitions, leads to the difference between Eqs.\,(\ref{eq:Rabi_DE}) and (\ref{eq:Rabi_KGE}). 

\begin{figure}[h]
\includegraphics[width=22pc]{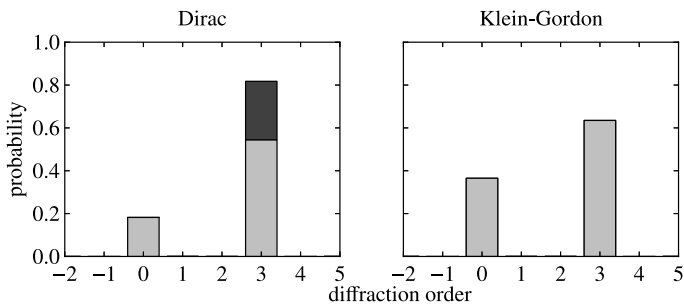}\hspace{2pc}%
\begin{minipage}[b]{14pc}\caption{\label{spec} Diffraction probability after an interaction time of 520 laser cycles ($T\approx 0.7$\,fs) for Dirac and Klein-Gordon particles. In the Dirac case, the light (dark) gray bars represent the spin-up (spin-down) probabilities. The other interaction parameters are given in the text.}
\end{minipage}
\end{figure}

The different Rabi frequencies $\Omega_{R,{\rm D}}$ and $\Omega_{R,{\rm KG}}$ give rise to different diffraction patterns for fermionic versus bosonic particles, as Fig.~\ref{spec} shows. The fact that the electron carries spin, enhances the Rabi frequency such that, after an interaction time corresponding to 520 laser cycles, the scattering probability is significantly larger than if the electron was spinless ($|c_{3,{\rm D}}(T)|^2\approx 80\%$ versus $|c_{3,{\rm KG}}(T)|^2\approx 60\%$). The left panel of Fig.~\ref{spec} also shows the contributions to the scattered state stemming from the spin-preserving and the spin-flipping transition. Note that the scattering probability of a Klein-Gordon particle differs not only from the spin-summed scattering probability in the Dirac case but also from the scattering probability without spin flip. This is because the interaction time is rather large, $T\sim\Omega_R^{-1}$. In contrast, when $T\ll\Omega_R^{-1}$ so that $\sin^2(\Omega_R T/2)\approx (\Omega_R T/2)^2$, 
the scattering probability of a Klein-Gordon particle would coincide with the spin-preserving scattering probability of a Dirac particle.

Hence, under certain conditions, the electronic spin degree of freedom can have a substantial impact on observable features of the three-photon Kapitza-Dirac effect.

\section{Spin dependence of photo-induced pair production }

Electron-positron pair creation can occur in a variety of physical contexts. In this section we focus ourselves on a generalized nonlinear version of the Breit-Wheeler reaction. In the generalized process an incident high-energy photon decays into a pair with simultaneous absorption of photons from a strong laser field. It has been studied first in the early 1960s \cite{Reiss,NR} and was observed experimentally in 1997 \cite{SLAC}. Spin-resolved calculations were provided in \cite{Tsai,SerboPP}. For very recent theoretical investigations of the process, we refer the reader to \cite{Kaempfer,KasiaBW}.

A comparative analysis between the photo-production rates $\mathcal{R}^{(0)}$ and $\mathcal{R}^{(\frac{1}{2})}$ of a pair of spin-zero and spin-$\frac{1}{2}$ particles, respectively, also provides evidence for the role played by the  internal spin degrees of freedom \cite{Selym}. The respective  calculations  of  the production rates can be based either on the quantum mechanical transition amplitude between Volkov states or, alternatively, on a consideration of the unitarity condition of the dispersion $S$-matrix. This latter procedure  provides  the  optical theorem, whose application to the case under consideration  allows us to establish a relation between the photo-production rate and the imaginary part of the vacuum polarization tensor $\Pi_{\mu\nu}^{(0,\frac{1}{2})}$. With the incident $\gamma$-photon energy $\omega_\gamma$, it reads
\begin{equation}\label{paircreapro}
\mathcal{R}^{(0,\frac{1}{2})}=\sum_{\ell=1,2}\frac{\epsilon_\ell^\mu\epsilon_\ell^{*\nu}}{2\omega_\gamma}{\rm Im}\ \Pi_{\mu\nu}^{(0,\frac{1}{2})}
=-\frac{1}{\omega_\gamma}{\rm Im}\left[ {\rm Tr}\ \Pi_{\mu\nu}^{(0,\frac{1}{2})}\right].
\end{equation} 
It is worth noting at this point  that the rates $\mathcal{R}^{(0,\frac{1}{2})}$ refer to the average over the polarization vectors $\epsilon_\ell^{\mu}$ of the incident high-energy photon beam whose intensity is much smaller than the intensity associated with the field of the strong wave. Observe, in addition, that the second step in the expression above comes out as a result of using  the  completeness  relation  $g^{\mu\nu}=-\sum_{\ell}\epsilon_\ell^\mu\epsilon^{*\nu}_{\ell}.$  Because of this fact, the photo-production rate of a pair of scalar (or fermionic) particles  is  finally expressed  in terms of the trace  ($\rm Tr$) of the corresponding polarization  tensor.  The  latter quantity  can be obtained from the general expression of $\Pi_{\mu\nu}$  in the field of a circularly  polarized  monochromatic plane-wave which was derived in \cite{baier}. As a consequence, 
\begin{eqnarray}
\label{scalarproba}
&&\mathcal{R}^{\left(0\right)}(\xi,\lambda)=-\frac{\alpha m^2}{2\pi\omega_\gamma}{\rm Im}\ \int_0^1dv\int_0^\infty \frac{d\rho}{\rho}e^{-\frac{2i\eta}{1-v^2}}
\left\{1-e^{iy}+2\xi^2\sin^2(\rho)\right\},\\
\label{spinorproba}
&&\mathcal{R}^{\left(\frac{1}{2}\right)}(\xi,\lambda)=\frac{\alpha m^2}{\pi\omega_\gamma}\mathrm{Im}\ \int_0^1dv\int_0^\infty \frac{d\rho}{\rho}e^{-\frac{2i\eta}{1-v^2}}\left\{1-e^{iy}-2\xi^2\frac{1+v^2}{1-v^2}\sin^2(\rho)\right\},
\end{eqnarray}
where $\alpha$ is the finestructure constant and the abbreviations  $\eta=(\rho/\lambda)\left[1+\xi^2\left(1-\sin^2(\rho)/\rho^2\right)\right]$ and $y=\frac{2\xi^2}{1-v^2} \frac{\rho}{\lambda}\left(1-\sin^2(\rho)/\rho^2\right)$ have been introduced, with $\lambda=k_\gamma\cdot k/(2m^2)$. Here, $k_\gamma=(\omega_\gamma,\vec{k}_\gamma)$ and $k=(\omega,\vec{k})$ are the four-momenta of the $\gamma$-photon and the laser photons, respectively. The remaining parameter of the  theory is  
\begin{eqnarray}
\xi=\frac{e|\vec{a}|}{m}\ ,
\label{xi}
\end{eqnarray}
with the amplitude $|\vec{a}|$ of the laser vector potential which is taken in radiation gauge. 

Unfortunately, an analytical  evaluation of  $\mathcal{R}^{(0,\frac{1}{2})}$ is quite difficult to perform in full glory.  However, the dependence on the parameters $\lambda$  and $\xi$ allows us to obtain closed-form expressions for the rates  in various asymptotic regimes of interest.  Let us start by  considering the case in which $\xi<1$ and $\lambda\gg 1+\xi^2$. Then, the scalar and fermionic pair production rates are given by
\begin{eqnarray}
\begin{array}{c}
\mathcal{R}^{(0)}\simeq\frac{\alpha m^2\xi^2}{4\omega_\gamma}\left[\left(1+\frac{n_*}{1+\xi^2}\right)\sqrt{1-n_*}-\frac{n_*}{1+\xi^2}\left(1-\frac{1}{2}n_* \right)\ln\left(\frac{1+\sqrt{1-n_*}}{1-\sqrt{1-n_*}}\right)\right]\theta(1-n_*),\\ \\
\mathcal{R}^{(\frac{1}{2})}\simeq\frac{\alpha m^2\xi^2}{2\omega_\gamma}\left[\left(1+\frac{n_*}{1+\xi^2}-\frac{n_*^2}{2(1+\xi^2)}\right)\ln\left(\frac{1+\sqrt{1-n_*}}{1-\sqrt{1-n_*}}\right)-\left(1+\frac{n_*}{1+\xi^2}\right)\sqrt{1-n_*}\right]\theta(1-n_*).
\end{array}\label{fermionprobabornsolu}
\end{eqnarray}
Some comments are in order. First of all, $n_*=2m^2(1+\xi^2)/k_\gamma\cdot k$ denotes  the minimal number of laser photons needed to overcome the energy threshold in the production process.   Note that  the  presence of the unit step function  $\theta(1-n_*)$ indicates that the reaction takes place with the absorption of one photon from the strong wave. We emphasize that the previous expressions   apply to a wider region with respect to the parameter 
$\xi$ and include, as a particular case,  the first Born approximation $(\xi\ll1)$.  The explicit expressions associated  with  this limit can be 
read off from  (\ref{fermionprobabornsolu})  by replacing $n_*\to 1/\lambda$ and  setting $\xi=0$ within the squared brackets. 

\begin{figure}
\includegraphics[width=0.47\textwidth]{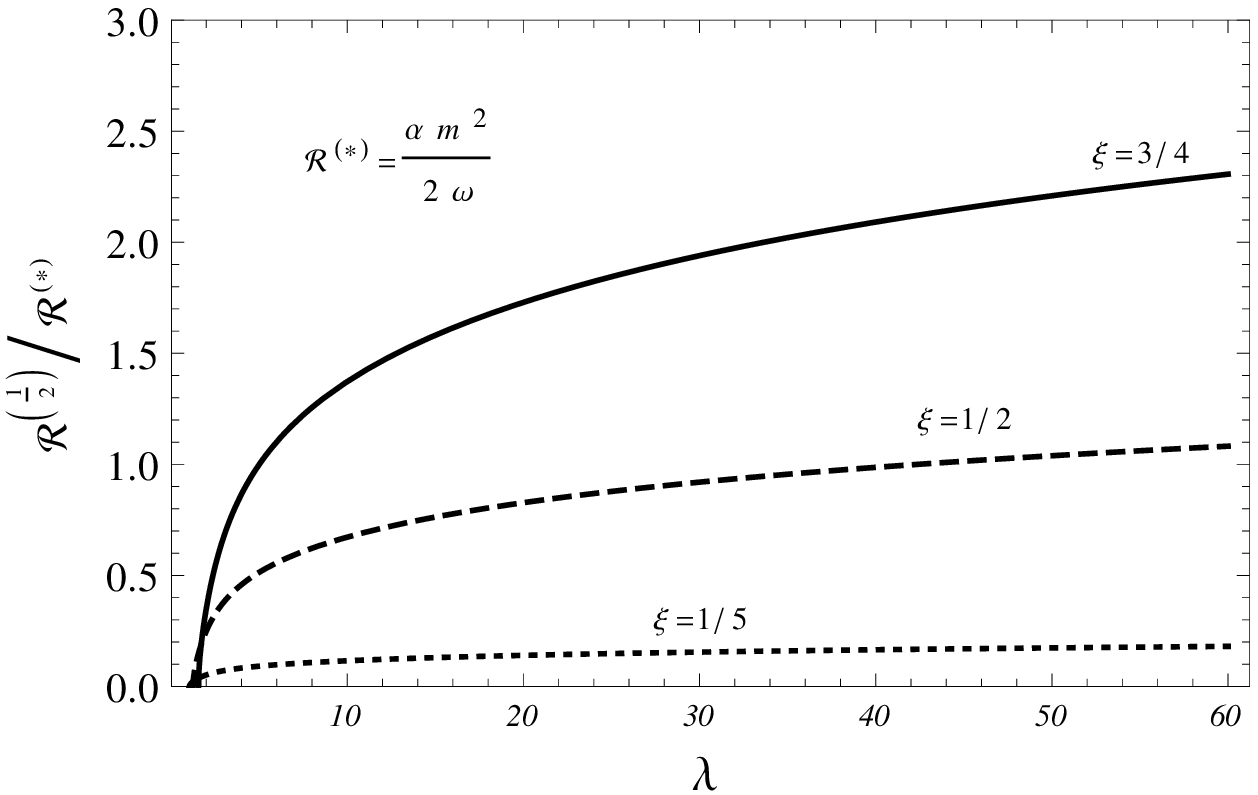}
\parbox[b]{0.52\textwidth}{
\bigskip
\medskip
\includegraphics[width=0.47\textwidth]{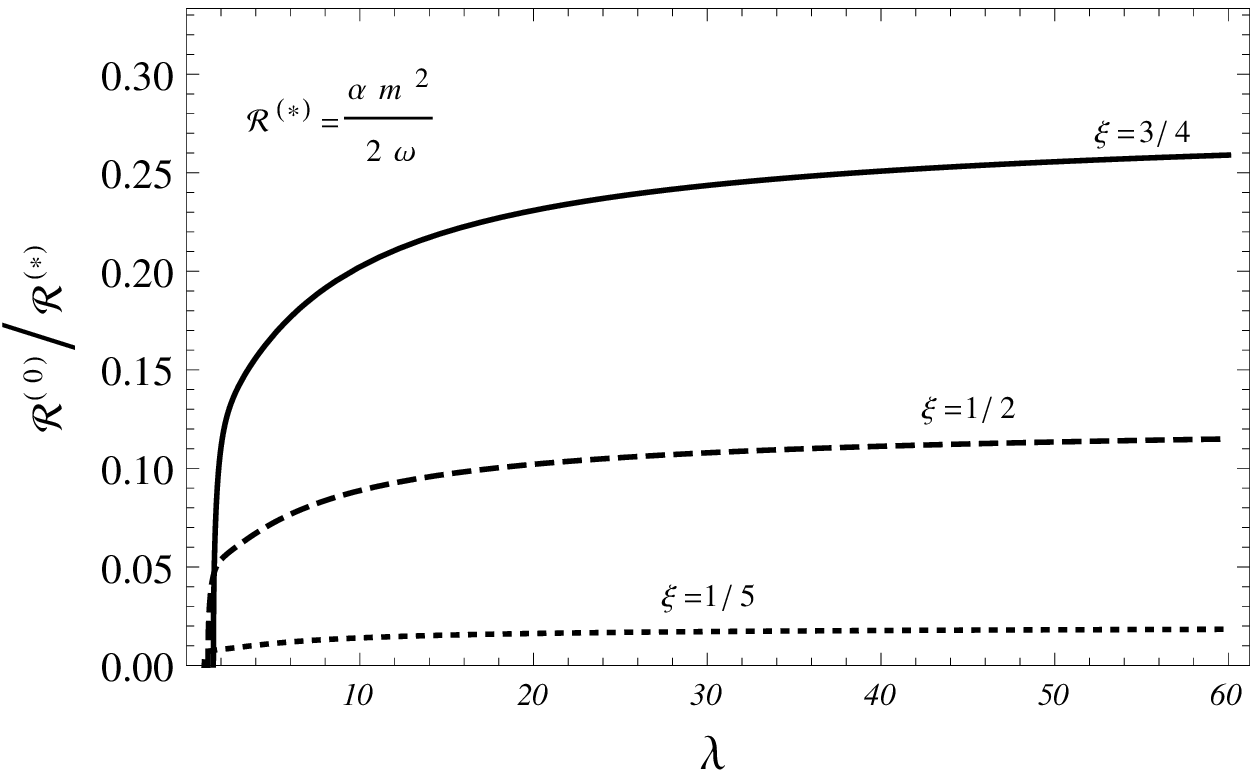}}
\caption{Dependence of the photo-production rates  of a pair of fermionic particles (left panel) and scalar particles (right panel) on the parameter $\lambda$ for different values of $\xi<1.$
\label{fig:rates}}
\end{figure}

The behaviours of   $\mathcal{R}^{(0)}$  and  $\mathcal{R}^{(\frac{1}{2})}$ under the asymptotic condition studied so far are displayed  in  Fig.~\ref{fig:rates}  as  a function of $\lambda.$ Several curves are shown corresponding to different values of $\xi.$  We point out that, for a common value of $\xi,$  the rate based on the Dirac theory is  generally found to be  larger than the corresponding results for Klein-Gordon particles,   i.e., $\mathcal{R}^{(\frac{1}{2})}>\mathcal{R}^{(0)}$.   

Consideration of the asymptotic regime $\xi\gg1$ is also of interest. This corresponds to the situation in which a large number of photons of the strong wave are absorbed (multiphoton reaction). In this case, the fermionic and scalar rates are  parametrized by a unique quantity  $\zeta=\lambda\xi$ which is independent of the frequency of the strong wave (quasistatic limit):
\begin{eqnarray}
\label{scalastrong}
\mathcal{R}^{(0)}&\simeq&\frac{\alpha m^2}{12\sqrt{3} \pi\omega_\gamma}\int_1^{\infty}\frac{du}{u\sqrt{u(u-1)}} K_{\frac{2}{3}}\left(\frac{4u}{3\zeta}\right)(4u-1)\,,\\
\mathcal{R}^{(\frac{1}{2})}&\simeq&\frac{\alpha m^2}{6\sqrt{3} \pi\omega_\gamma}\int_1^{\infty}\frac{du}{u\sqrt{u(u-1)}} K_{\frac{2}{3}}\left(\frac{4u}{3\zeta}\right)(8u+1)
\end{eqnarray}
where $K_{\frac{2}{3}}(x)$ denotes the Macdonald function. Using the behaviour of this function for small and large arguments, respectively, one obtains
\begin{eqnarray}
&&\mathcal{R}^{(0)}\simeq\frac{\alpha m^2 \zeta^{\frac{2}{3}}}{6\sqrt{3\pi}\omega_\gamma}\left(\frac{3}{2}\right)^{\frac{2}{3}}\frac{\Gamma^2\left(\frac{2}{3}\right)}{\Gamma\left(\frac{13}{6}\right)}\ , \qquad \mathcal{R}^{(\frac{1}{2})}\simeq\frac{5\alpha m^2 \zeta^{\frac{2}{3}}}{6\sqrt{3\pi}\omega_\gamma}\left(\frac{3}{2}\right)^{\frac{2}{3}}\frac{\Gamma^2\left(\frac{2}{3}\right)}{\Gamma\left(\frac{13}{6}\right)} \quad \mathrm{for}\quad \zeta\gg1\\
&&\mathcal{R}^{(0)}\simeq\frac{\alpha m^2 \zeta}{16\omega_\gamma}\sqrt{\frac{3}{2}}\,e^{-\frac{4}{3\zeta}}\ , \ \ \ \qquad\quad\quad \mathcal{R}^{(\frac{1}{2})}\simeq\frac{3\alpha m^2 \zeta}{8\omega_\gamma}\sqrt{\frac{3}{2}}\,e^{-\frac{4}{3\zeta}} \!\!\!\qquad\qquad\mathrm{for}\quad \zeta\ll1
\end{eqnarray}
with $\Gamma(x)$ denoting the Gamma function. It follows from the expression above that  $\mathcal{R}^{\left(\frac{1}{2}\right)}=5\mathcal{R}^{(0)}$ for $\zeta\gg1$, whereas $\mathcal{R}^{\left(\frac{1}{2}\right)}=6\mathcal{R}^{(0)}$ for $\zeta\ll1$. We point out that the production rates $\mathcal{R}^{\left(\frac{1}{2}\right)}$ for fermion pairs may also be found in the books \cite{books}.

The general conclusion associated with these results is that the spin degree of freedom is clearly manifest in the total pair production rates. In particular, the lack of spin in the case of scalar particles suppresses  the pair creation  rate  as compared to the predictions from Dirac theory. Thereby, the spin dependence is non-trivial because it differs from a simple factor of four which would account for the number of possible configurations
of two spin-$\frac{1}{2}$ particles.

\section{Spin-resolved multiphoton pair production on a nucleus}
Electron-positron pairs can also be produced in the combined fields of a strong laser beam and an atomic nucleus \cite{review1,review2} (for some very recent papers see \cite{KasiaBH,PiazzaPLB}). When the nucleus is counterpropagating the laser beam with a high Lorentz factor $\gamma$, then the relativistic Doppler shift of the laser photon energy and laser field strength can be exploited. The process may be considered as a generalization of the well-known Bethe-Heitler effect to the multiphoton case. 

Recently, spin effects in the multiphoton Bethe-Heitler process have been examined \cite{Tim-Oliver1,Tim-Oliver2,Piazza}. These studies were motivated by the question whether high degrees of spin polarization can be obtained when particle pairs are produced via the absorption of many photons. So far, the asymptotic regimes with $\xi\ll 1$ and $\xi\gg 1$ have been considered [see Eq.\,(\ref{xi})]. Here we study the intermediate regime characterized by a laser intensity parameter of order unity ($\xi\sim 1$).

Within the strong-field approximation, the amplitude for multiphoton Bethe-Heitler pair production may be written as
\begin{eqnarray}
\label{S}
{\mathcal M}_{p_+s_+,p_-s_-} = -ie\int d^4x\,[\Psi_{p_-,s_-}^{(-)}]^\dagger V(r)\Psi_{p_+,s_+}^{(+)}.
\end{eqnarray}
The electron and positron are described by Dirac-Volkov states $\Psi_{p_\pm,s_\pm}^{(\pm)}$ which are labeled by the free four-momenta $p_\pm=({E_\pm,\vec p}_\pm)$ and spin quantum numbers $s_\pm$ outside the field. The laser field is assumed to be a circularly polarized plane wave. Below we shall consider the spin basis of helicity states, describing right- and left-handed particles, respectively. $V(r)=Ze/r$ denotes the Coulomb potential of the nucleus of charge number $Z$ in its rest frame; it is taken into account within the lowest order of perturbation theory.

From the amplitude (\ref{S}) one can derive the pair production rate in the usual manner. First, by expanding the periodic parts in the integrand of Eq.\,(\ref{S}) into a Fourier series, one can decompose the amplitude into the contributions stemming from absorption of a certain photon number $n$,  according to ${\mathcal M}=\sum_n {\mathcal M}^{(n)}$. By taking the absolute square of the amplitude, integrating over the degrees of freedom of the produced electron, and summing over $n$ one obtains the double-differential rate $d^2 {\mathcal R}^{\rm{R},\rm{L}}/dE_+ d\vartheta_+$ for detecting a positron with a certain emission angle $\vartheta_+$ and energy $E_+$ and either positive $s_+=s_{\rm R}$ or negative $s_+=s_{\rm L}$ helicity \cite{Tim-Oliver2}. The asymmetry ratio between the rates for measuring right-handed and left-handed positrons,
\begin{equation}
 P_\parallel(E_+,\vartheta_+) = \frac{d^2 {\mathcal R}_{\rm R} - d^2 {\mathcal R}_{\rm L}}{d^2 {\mathcal R}_{\rm R} + d^2 {\mathcal R}_{\rm L}}\,,
\label{eq:degree_2}
\end{equation}
determines the degree of longitudinal polarization. We point out that, within the present approach, the degree of longitudinal polarization for the electron will exhibit the same energy and angle dependence.

\begin{figure}[b]
\includegraphics[width=22pc]{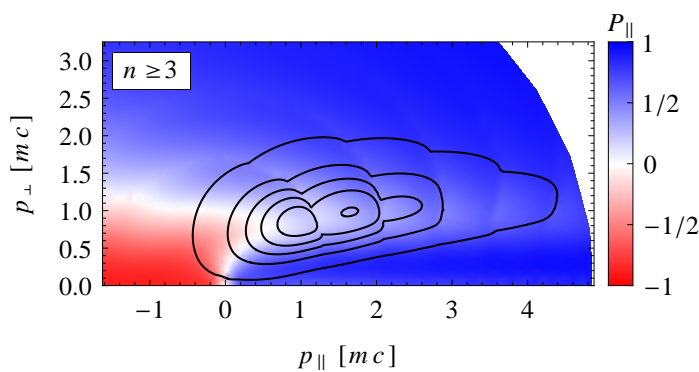}\hspace{2pc}%
\begin{minipage}[b]{14pc}\caption{\label{DoP} Color-coded degree of longitudinal polarization [see Eq.\,(\ref{eq:degree_2})] as a function of the transverse and parallel momentum components of the positron with respect to the laser beam (including positron energies up to $5mc^2$). Regions of negative (positive) degrees of polarization are encoded red (blue). Contours of the double-differential production rate are plotted as black lines.}
\end{minipage}
\end{figure}

Figure \ref{DoP} shows the degree of longitudinal polarization of the positrons produced via the multiphoton Bethe-Heitler effect by an intense laser field with $\xi=0.5$. The results refer to the nuclear rest frame. In this frame, the laser photons have an energy of $\omega_{\rm nuc} = 0.4$\,MeV such that electron-positron pairs can be produced by absorption of three or more photons. As can be seen from the figure, most of the momentum configurations support high positive degrees of polarization. Negative degrees of polarization only appear for particles that are emitted roughly antiparallel to the laser beam with relatively low energies. However, for practical purposes it is relevant to correlate the degree of polarization obtainable for a certain positron momentum vector with the corresponding rate of production.

Therefore, in addition to the color-coded degree of polarization, Fig.\,\ref{DoP} also contains contour lines indicating the absolute value of the double-differential production rate. They are plotted at every fifth part of the maximum rate. Note that the contributions of the various photon orders are clearly visible in the contour lines. For the present parameters, the dominating photon orders are $n = 4,\ldots, 8$ whose relative contributions to the total production rate are 23\%, 34\%, 23\%, 12\% and 5\%, respectively, whereas the relative contribution of the lowest energetically allowed order $n=3$ is very small ($\sim 3.5 \times 10^{-4}$) due to a channel-closing effect. The majority of particles is produced inside the two innermost contour lines, where only moderate degrees of polarization are achieved, as the color-coding illustrates.

The parameter regime just discussed in the nuclear rest frame can be realized, for instance, by using a high Lorentz factor of the nucleus of $\gamma =2000$ in the laboratory frame and an extreme ultraviolet laser photon energy of 100\,eV. With an invariant intensity parameter of $\xi = 0.5$ this corresponds to a lab-frame laser intensity of $I \approx 4.4 \times 10^{21}$\,W/cm$^2$. 

\begin{figure}[h]
\includegraphics[width=22pc]{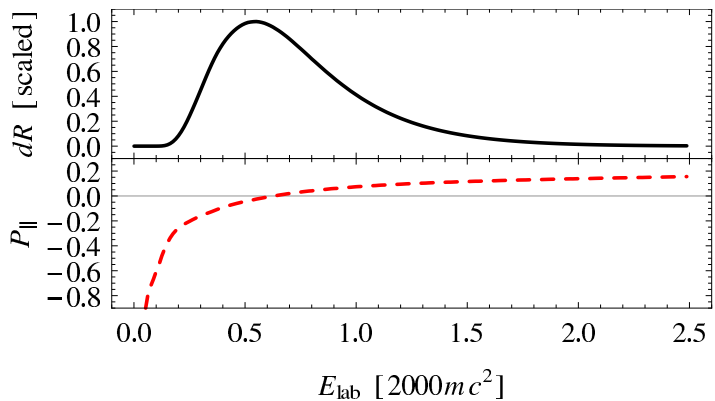}\hspace{2pc}%
\begin{minipage}[b]{14pc}\caption{\label{label} Scaled production rate, differential with respect to the positron energy (black solid line), together with the corresponding degree of longitudinal polarization (red dashed line) in the laboratory frame. The parameters of the laser-nucleus collision are 
$\omega_{\rm lab} = 100$\,eV, $\xi = 0.5$ (i.e. $I \approx 4.4 \times 10^{21}$\,W/cm$^2$) and $\gamma = 2000$. At the maximum of the differential rate, the degree of polarization is very small.}
\end{minipage}
\end{figure}

Figure~\ref{label} shows the energy differential production rate in the laboratory frame in scaled units (black solid line). Note that, in contrast to the results from the nuclear rest frame, the production rates stemming from different photon orders cannot be resolved, which is a result of the Lorentz transformation. The maximum of the spin-summed production rates occurs at a positron energy of around $0.6 \gamma mc^2$. As the red dashed line in Fig.~\ref{label} shows, this is where the degree of longitudinal polarization is just the lowest. For those energies where nonnegligible rates are attained, the average degree of longitudinal polarization \cite{average} is bounded to values between about $\pm 0.2$. 

In conclusion, only relatively low degrees of longitudinal polarization can be obtained via the multiphoton Bethe-Heitler process in the intermediate intensity regime. This is in agreement with the previous results for $\xi\gg 1$ and $\xi\ll 1$ in \cite{Tim-Oliver1,Tim-Oliver2}. It is also in accordance with the result of \cite{Piazza} where it was shown that, in the quasistatic regime, the spin polarization vector of the positron is directed, to leading order, along the transverse momentum component. Hence, although the pairs are produced by absorption of several laser photons which all carry the same helicity, the transfer of helicity is less efficient than in the traditional Bethe-Heitler process involving one photon of high energy \cite{McVoy}.

\section{Conclusion}
We studied the influence of the electron spin on three different fundamental quantum processes which can occur in the presence of strong laser fields. It was shown that, as one may have expected, the electron spin plays an important role in pair production processes such as the multiphoton Breit-Wheeler and Bethe-Heitler reactions. Besides, under certain conditions, the electron spin can also have a characteristic impact on Kapitza-Dirac scattering from a standing wave of light where it manifests itself in the Rabi oscillation period and the diffraction pattern. 

We point out that with the present contribution our previous articles \cite{Sven,Selym,Tim-Oliver1,Tim-Oliver2,Piazza} have been extended by examining spin effects in electron-positron pair creation by the multiphoton Bethe-Heitler effect in an intermediate coupling regime ($\xi\sim 1$), which has not been considered before. Besides, we  explicitly showed the Rabi oscillation dynamics of a Klein-Gordon particle scattering from a standing laser wave and the photo-production rates of Dirac versus Klein-Gordon particles at $\xi<1$.   

As a general result we note that, in all three cases under consideration, the process probabilities for Dirac particles were found to be higher than for Klein-Gordon particles.

\section*{Acknowledgments}
S.~V.-C. gratefully acknowledges support from the Alexander von Humboldt Foundation.


\begin{thebibliography}{9}
\bibitem{negative} The amplitudes $c_3^{\zeta}(T)$ with $\zeta \in \mbox{$\{ {-}\uparrow, {-}\downarrow \}$}$ corresponding to negative electron energies are practically zero.
\bibitem{average} The longitudinal degree of polarization shown in Fig.\,\ref{label} involves an average over the positron emission angle which varies within very narrow limits in the laboratory frame due to the large value of the Lorentz factor.
\end{thebibliography}
\end{document}